\def \myplotone#1 {\plotone{#1}}
\def \myplotfiddle#1 {\plotfiddle{#1}{512pt}{0}{100}{100}{-288}{-140}}

\def\myplottwo#1#2 {\plottwo{#1}{#2}}

\def \etal {{\it et al. }}
\def \Mpc {h^{-1}{\rm Mpc}}
\def \kpc {h^{-1}{\rm kpc}}
\def \farcs{\hbox{$.\!\!^{\prime\prime}$}}
\def \farcm{\hbox{$.\!\!^{\prime}$}}

\def \Vcomplim {24.0}
\def \Icomplim {23.5}
\def \NobjI    {1828}
\def \NobjV    {1255}
\def \NgalI    {922}

\documentstyle[aaspp4,12pt]{article}

\eqsecnum
\slugcomment{Submitted to Ap.J.}

\begin{document}

\title{A Weak Gravitational Lensing Analysis of Abell~2390}
\author{G. Squires\altaffilmark{1}
N. Kaiser\altaffilmark{2}\altaffilmark{,5}, 
G. Fahlman\altaffilmark{3}\altaffilmark{,5}, A. Babul\altaffilmark{4} and
D. Woods\altaffilmark{3}}
\altaffiltext{1}{Center for Particle Astrophysics, University of 
California, Berkeley, CA 94720 USA}
\altaffiltext{2}{Canadian Institute for Advanced Research and
Canadian Institute for Theoretical Astrophysics, University of Toronto,
60 St.\ George St., Toronto, Ontario, Canada M5S 1A7}
\altaffiltext{3}{Department of Geophysics and Astronomy, University of
British Columbia, 2219, Main Mall, Vancouver, BC, Canada V6T 1Z4}
\altaffiltext{4}{Department of Physics, New York University,
4 Washington Place, Room 525, New York, New York, USA 1003-6621}
\altaffiltext{5}{Visiting Astronomer, Canada-France Hawai'i Telescope.
Operated by the: National Research Council of Canada, le Centre National
de la Recherche Scientifique de France and the University of Hawai'i}

\begin{abstract}
We report on the detection of dark matter in the cluster Abell~2390 
using the weak gravitational distortion of background galaxies. 
We find that the cluster light and total mass distributions are
quite similar over an angular scale of $\simeq 7^\prime \;(1 \Mpc$).
The cluster galaxy and mass distributions are centered on the cluster
cD galaxy and exhibit elliptical isocontours in the central 
$\simeq 2^\prime \; (280 \kpc)$.
The major axis of the ellipticity is aligned with the direction defined
by the cluster cD and a ``straight arc'' located $\simeq 38^{\prime\prime}$
to the northwest.
We determined the radial mass-to-light profile for this cluster and found
a constant value of  $(320 \pm 90) h\; M_\odot/L_{\odot V}$, which is
consistent with other published determinations. We also compared our 
weak lensing azimuthally averaged radial mass profile with a spherical 
mass model proposed by the CNOC group on the basis of their detailed
dynamical study of the cluster. We find good agreement between the two
profiles, although there are weak indications that the CNOC density
profile may be falling more steeply for $\theta\geq 3^\prime$ $(420\kpc)$.
\end{abstract}


\section{Introduction}
Understanding the nature of the mass distribution in clusters of
galaxies is a key cosmological issue. Typically, three independent
methods to probe the cluster mass distribution have been
applied: 1) applications
of the virial theorem to the cluster galaxy velocity distribution, 2)
analyses of the X-ray emission of the diffuse, intracluster gas and
3) mass determinations based on the gravitational lens distortion of
background objects by the foreground cluster.  The latter technique,
the analyses of both the strong as well as the weak gravitationally induced 
distortions in the images of faint background galaxies, is 
particularly attractive as it does not require any assumptions about
the geometry or the dynamical of the clusters, with the weak 
distortions being especially suited for mapping the cluster mass 
distribution out to large radii.

Over the past few years, there has been considerable
activity to develop a theoretical framework for determining the
mass distribution in the cluster based on a measured set of
gravitationally distorted galaxy shapes
(eg. \cite{tyson90}; \cite{ks93}; \cite{schneider94}; 
\cite{kaiser95b}; \cite{schneider95}; 
\cite{seitzc95};  \cite{seitzs95}, \cite{sk95}).
Several groups have developed algorithms and software designed to measure
and analyze very small distortions expected in the cluster outskirts 
(\cite{ksb95}; \cite{bonnet95}; \cite{fischer96}).
These techniques have been applied
to the central $\simeq 1 \Mpc$ of several galaxy clusters (eg. \cite{tyson90};
\cite{fahlman94}; \cite{bonnet94}; \cite{tyson95}; \cite{squires95}).

For a few clusters, the results of the weak lensing analyses has compared with 
the results of X-ray analyses as well as detailed spectroscopic studies.  
There is no {\it apriori} reason to expect an agreement between the different
analyses and in fact, such joint analyses can potentially enable to us
learn much more about the relative distributions and the dynamical 
states of the gas, galaxies and dark matter in the clusters than is
currently known.  Indeed, in the central regions of some clusters 
(eg. A2218 and A1689), the mass inferred 
{}from the presence of (strong) lensing
features exceeds that derived {}from the X-ray data under the standard
assumption of pressure supported hydrostatic equilibrium by a factor
of $\simeq 2$--$3$ (\cite{meb95}).  A similar discrepancy also arises
in galaxy cluster MS~1224, where the weak lensing inferred mass
is $\simeq 2$ times larger than 
the mass determined {}from a virial analysis of the 
cluster galaxies (\cite{fahlman94}; \cite{carlberg94}).
Also, in a joint lensing/X-ray analysis of A2218  (\cite{squires95}),
the lensing mass profile was found to be systematically a factor of 
$\simeq 2$ larger than the X-ray mass profile, although the two are 
formally in agreement within the 95\% confidence level. 
A recent analysis based on ASCA data confirmed this discrepancy
(\cite{loewenstein96}).
Conversely, a
similar analysis of the cluster A2163 yielded good agreement between
the weak lensing and X-ray mass estimates (\cite{squires96}).

In this paper, we discuss the dark matter and galaxy distributions
in the cluster of galaxies Abell~2390. 
Abell 2390 is an Abell richness class 1 cluster (\cite{abell89})
at redshift $z = 0.2279$ with a measured line of sight velocity dispersion 
$\sigma = 1093$~km/s (\cite{carlberg95}). 
Abell~2390 has an X-ray luminosity of
$L_x = 9.77 \times 10^{44}$~ergs/s (H~=~75~km/s/Mpc) in the 2-6~keV
energy range (\cite{kowalski84}) and $L_x = 4.67 \times 10^{44}$~ergs/s in 
the 0.7-3.5~keV range (\cite{ulmer86}).
A2390 has been the focus of several studies 
(eg. \cite{pello91}, \cite{kassiola92}, \cite{narasimha93}) seeking to 
account for the observed lensing features such as the ``straight arc''
located approximately 38$^{\prime\prime}$ away {}from the central galaxy 
(\cite{pello91}).

In this analysis, the total mass distribution in A2390 is 
determined using the weak 
gravitational distortion of faint background images induced by the cluster.  
The data reduction and analysis procedure used here is very similar
to that we employed in our analysis of A2218 (\cite{squires95}) 
and A2163 (\cite{squires96}) and we
refer to those papers for further detail. Here we will concentrate mainly
on the results.  
We also compare our results for the morphology of the  mass distribution 
and the estimate for the  total mass with published results {}from galaxy 
spectroscopy and X-ray analyses. 

\section{Data Acquisition and Analysis}

The cluster was observed at the 3.6m Canada-France-Hawai'i telescope
on the nights of 1994 June 6-9. The detector used was the 2048~x~2048
Loral 3 CCD at prime focus with a pixel size of 0\farcs207.  
For the cluster Abell~2390, we obtained $5 \times 15$~minute exposures in the 
I-band and $2 \times 30$~minute exposures in V with seeing conditions of
FWHM=0\farcs6 and 0\farcs8 in I and V respectively. 
We observed the central 7$^\prime$ of the cluster,
covering a square of side $\simeq 1\Mpc$ at the cluster redshift. 

A detailed discussion of the observations and data reduction are described 
in \cite{squires95} . Very briefly, each bias subtracted image was divided 
by a median twilight flat. The residual rms scatter in the sky background on
the individual images was 23.9 magnitudes per 
square arcsecond in I, and 25.3 magnitudes per square arcsecond in V.
The data was calibrated against photometric standards in the globular
clusters M92 and NGC 4147 (unpublished photometry {}from Davis 1990;
see also Stetson and Harris 1988 and Odewahn \etal 1992) and 
Landolt (1992) standards in SA110. Color terms were found to be
unnecessary in the transformation and the I and V zero points were
determined with a formal error of less than 0.005 mag.

We detected objects with a significance
of $4\sigma$ over the local sky background in each frame and selected
objects that were detected on least two of the images in each waveband
independently.
We formed master catalogues
of $\NobjI$ objects in I and $\NobjV$ in V which are
complete to $\simeq \Icomplim$ and $\simeq \Vcomplim$ 
in I and V respectively.
We measured the position, shape and brightness parameters for all of the
objects using our standard
procedure (\cite{ksb95}).

\section{Cluster Light Distribution}

Combining the V- and I-band observations, we determined the color of objects
detected in both wavebands. We identified a red sequence of objects 
with mean color of $V-I \simeq 1.7$ (uncorrected for reddening)
at the bright end which we classified 
as cluster galaxies. To extract this color selected cluster galaxy sample, 
we fitted a linear model to the 
color sequence and selected bright objects $(I<21)$ with color within 
$0.2$ magnitudes of the mean.

To obtain a qualitative description of the cluster light distribution,
we calculated the galaxy light and galaxy surface number density. We
employed the color selected sample and display as contour plots and  3D
surface plots
the cluster galaxy light distribution and surface number density
in Figure \ref{fig:a2390_lightandmass}. The
distributions have been smoothed with a Gaussian filter with scale
0\farcm5 (this scale
was chosen match the smoothing used in the lensing analysis).
The galaxy number and light distributions are very similar.
The dominant structure in both distributions is centered on the
giant elliptical galaxy and exhibits some ellipticity in
the central regions, with the major axis oriented in the direction 
towards the the straight arc located $\simeq 38^{\prime\prime}$ 
to the northwest of the central galaxy. There is a 
small secondary peak
approximately $4^\prime$
to the northwest of the cD galaxy that appears in both the number and
luminosity weighted plots. The height of the secondary peak is
$\simeq 20$\% of the amplitude of primary concentration in the smoothed light
map.

We estimated the total light content in the cluster in two ways.
First, we formed a lower bound on the cluster light by only including the 
light associated  with galaxies having the mean cluster galaxy color. This  
excludes cluster members with color outside the main cluster color sequence 
as well as galaxies with magnitudes beyond the faint-end limit.
Secondly, we calculated the light in the cluster by adding the
light associated with {\em all} the galaxies detected in the image. Since
this sample includes both cluster as well as field galaxies, the resulting
measure of the light provides an upper limit to the light content of the 
cluster, down to the limiting magnitude of the catalogue.
We calculated the zero redshift V-band luminosity by
$L_V = 10^{ 0.4 ( M_{V \odot} - V + DM + K + A_V) } L_{\odot V}$
where $M_{V \odot} = 4.83$ is the total solar V magnitude. We applied
a K-correction of $K = 0.5$
as suggested by Coleman, Wu and Weedman (1980).  Since A2390 
is at redshift $z=0.2279$, its distance modulus is $DM = 39.28$ (for 
$\Omega = 1$ and $h = 1$) and  
we used an extinction correction of $A_V = 0.25$.	

\begin{figure}
\myplotone{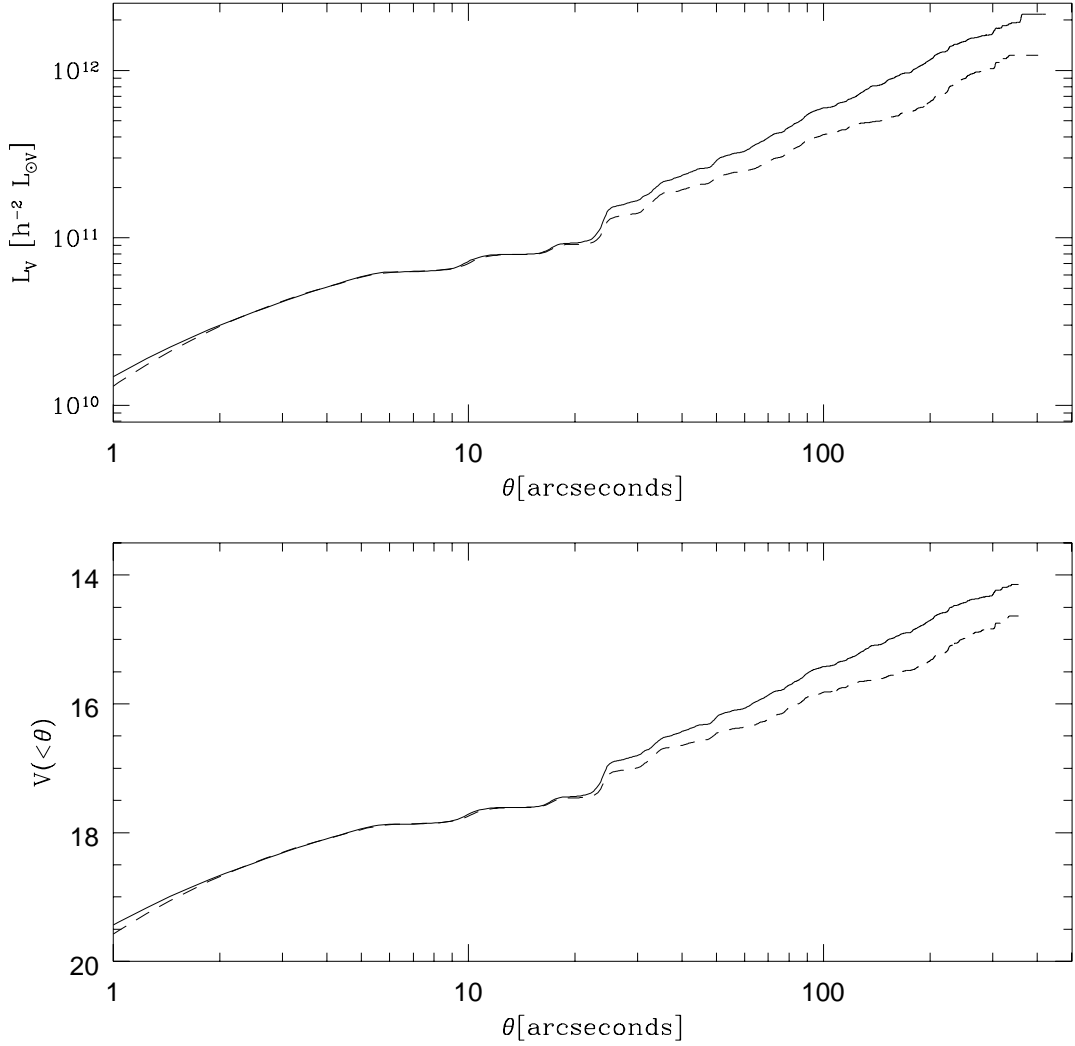}
\caption{The cumulated V-magnitude and $L_V$ luminosities {}from the central
elliptical galaxy. The solid lines
come {}from adding the light {}from all galaxy candidate objects in the
field. The dashed line is for the color selected sequence.}
\label{fig:LightProfiles}
\end{figure}

The cumulated light profiles are shown in
Figure \ref{fig:LightProfiles}. The solid line shows the results using all the
galaxies in the images, while the dashed line shows the light profile
computed {}from the color selected sample.	
The slope and amplitude of 
the light of the full galaxy sample exceeds that of the cluster 
selected subsample and for $\theta \geq 100^{\prime\prime}$, the former is
a factor of $\geq 1.5$ greater.  The true cluster light profile is likely to lie
in between the two estimates.  Regardless, in computing quantities such as
the cluster mass-to-light ratio, we adopt a conservative approach and use
the higher of the two profiles.

\section{Lensing Analysis}

We identified galaxies as objects having half-light radius greater than
1.2 times the mean stellar half-light radius and
applied a magnitude cut of $21 < I < 24$. The bright magnitude limit was
chosen, somewhat arbitrarily, to exclude cluster and foreground galaxies, but
the results are very insensitive to the actual threshold level.
With these cuts, we obtained $\NgalI$ galaxies for this analysis.
We present here the I-band data for the lensing analysis as the seeing
conditions were significantly better than the V-band observations
(although the V-band data was used as a consistency check). 
To convert the measured galaxy shapes into estimates of the cluster
surface density, we corrected for the effects of anisotropy
in the point spread function (psf) due to wind-shake, guiding errors, etc.,
and the smoothing due to seeing using the procedure described in
Kaiser, Squires \& Broadhurst (1995).

\begin{figure}
\myplotone{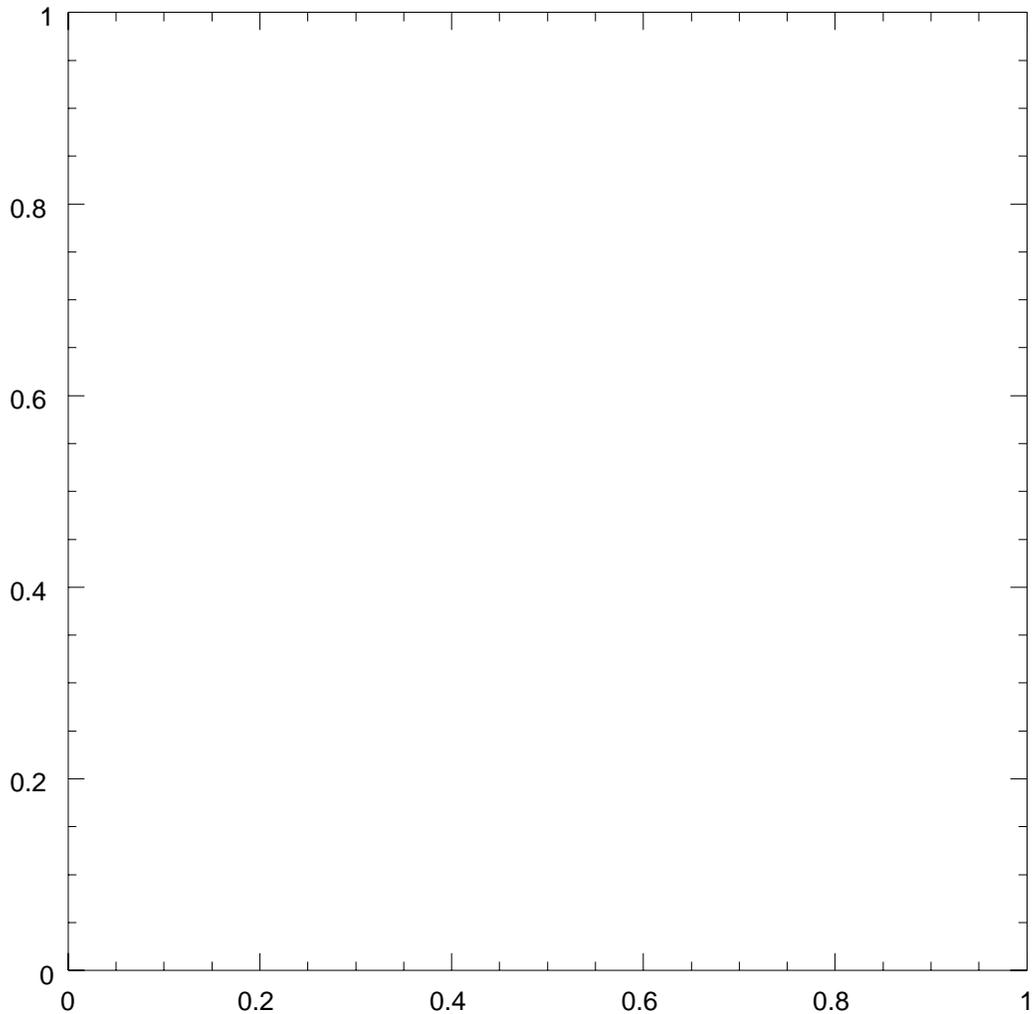}
\caption{The bottom two rows show the luminosity and number weighted 
distributions of the color selected 
cluster galaxy sample.
The top row is the reconstructed mass distribution.
The distributions have been smoothed
with a Gaussian filter with scale 0\farcm5. In the contour plots,
North is to the right; East is up. The box size is $7^\prime = 966 \kpc$.
This figure can be obtained at ftp://magicbean.berkeley.edu/pub/squires/a2390/massandlight.ps.gz}
\label{fig:a2390_lightandmass}
\end{figure}

We used the seeing and psf-corrected galaxy shape measurements to
determine the surface mass density in the cluster and display the
results in Figure \ref{fig:a2390_lightandmass}. The surface
density reconstruction was determined using
the maximum probability extension to the original Kaiser \& Squires (1993)
algorithm (\cite{sk95}). We employed 15 wavemodes and a regularization 
parameter of $\alpha = 0.05$ although we note that the results are quite 
insensitive 
to variations in $\alpha$ by a factors of $\simeq 5$.

The peak of the mass concentration coincides with the
central giant elliptical galaxy and is significant at the
$\simeq 6\sigma$ level.
For $\theta \geq 1^\prime$, 
the mass isocontours are elongated in the southeast--northwest direction,
in alignment with the direction defined by the central galaxy and the
straight arc.  There is an extension towards the north 
at a distance of $\simeq 2^\prime$ of the central galaxy which lies
in the direction of the overdensity that appeared in the
cluster number and light distributions. The peak itself (if it exists in
the mass distribution) is not
resolved. This is not unreasonable given
the number of galaxies used in this reconstruction: assuming that mass
scales directly with light, then we would only
expect to detect the secondary mass fluctuation at the $\simeq 1\sigma$
level. Thus, with the surface density of galaxies available here, we can not
map with high statistical significance the mass associated with 
small fluctuations in the cluster galaxy distribution.

To determine the total mass in the cluster, we employed the 
statistic (Kaiser \etal 1995)
\begin{eqnarray}
\zeta(\theta_1, \theta_2)& =& 2( 1 - \theta_1^2 / \theta_2^2 )^{-1}
	\int_{\theta_1}^{\theta_2} d \ln(\theta) \langle \gamma_t \rangle 
\end{eqnarray}
which measures the mean dimensionless surface density interior 
to $\theta_1$, relative
to the mean in an annulus $\theta_1 < \theta < \theta_2$. 
To convert $\zeta$ into a physical quantity, we estimated
$\Sigma_{crit} = (4\pi G D_l \beta)^{-1} $ where
$D_l$ is the angular diameter distance to the lens and
$\beta = \hbox{max}( 0, \langle 1 - w_l /w_s 
\rangle )$.  In an Einstein de-Sitter universe with $\Omega = 1$, 
the comoving distance $w$ is defined as $w = 1 - 1/\sqrt{1+z}$.
We estimated $\beta$ by an extrapolation of the faint redshift
surveys (\cite{lilly93}; \cite{tresse93})
to fainter magnitudes and find,
for A2390,  
$\Sigma_{crit} = (6.5 \pm 0.6) \times 10^{15} h M_\odot$/Mpc$^2$.

\begin{figure}
\myplottwo{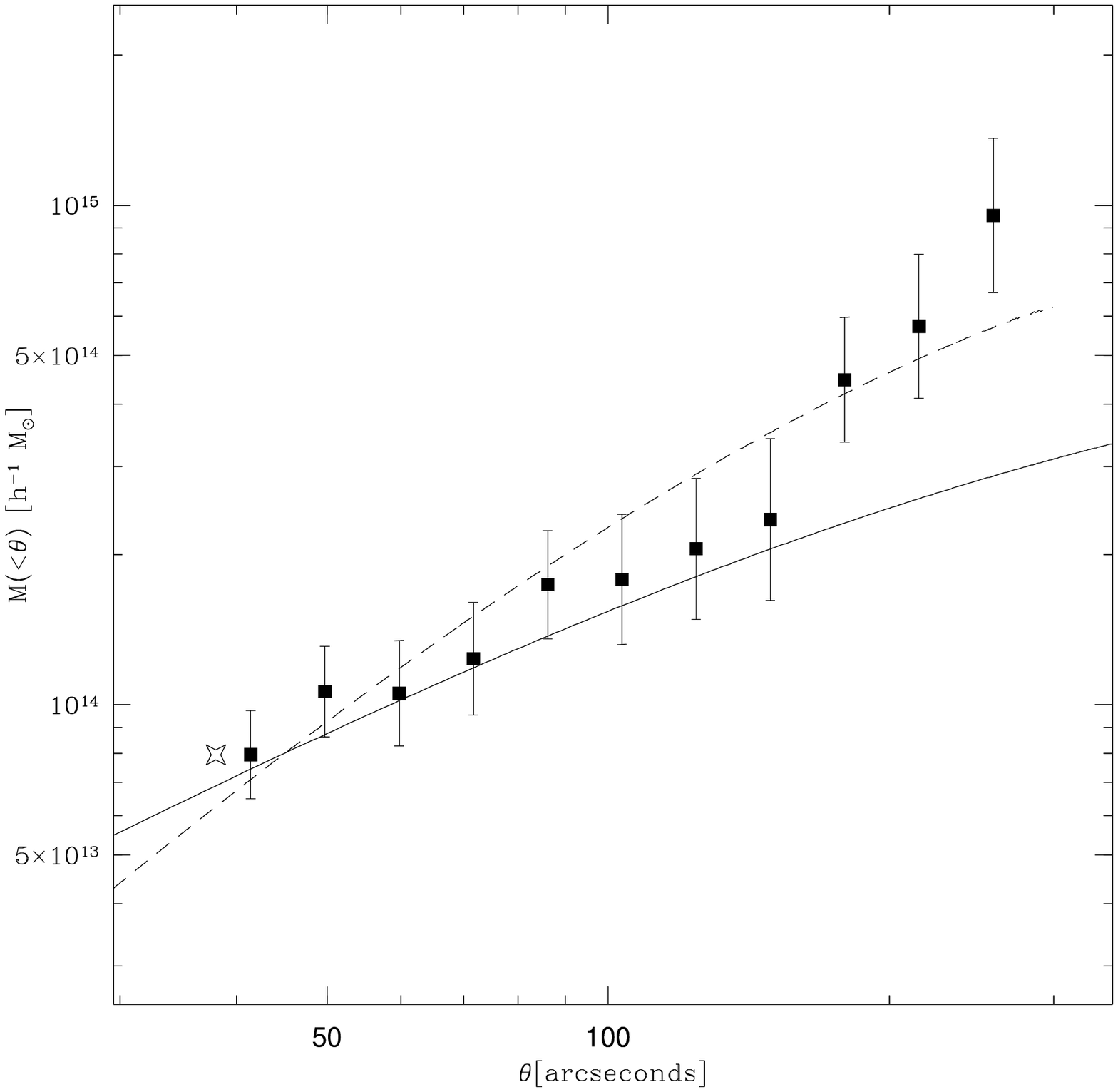}{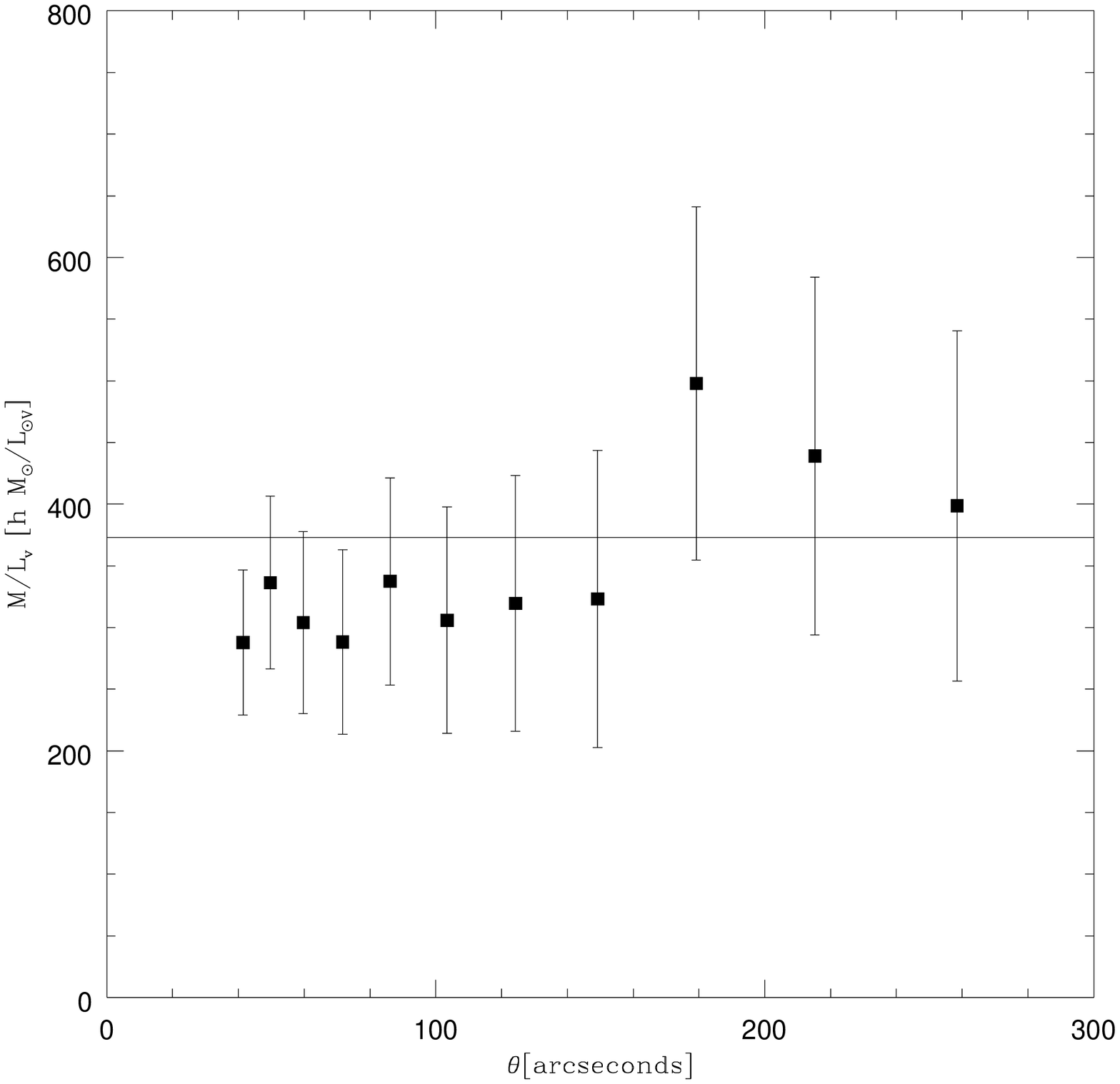}
\caption{Left panel: the radial mass profile. The solid line is the 
prediction for an isothermal model with the measured velocity dispersion 
of $\sigma = 1093$~km/s and the dashed line is the CNOC model 
(Carlberg \etal 1996). 
Both predictions have been corrected for matter in
the control annulus to facilitate direct comparison with the lensing
data. The ``x'' is
the mass estimated by Pierre \etal (1995) using a bimodal model.
Right panel: the radial mass-to-light profile,
determined by taking the ratio of the mean surface mass and light 
densities as a function of radius, relative to the mean in the control annulus.
}
\label{fig:a2390_MassvsR}
\end{figure}

Using the estimate of $\Sigma_{crit}$, we convert the $\zeta$ estimates
into a mass. In Figure \ref{fig:a2390_MassvsR} we display the radial mass
profile of Abell~2390 and the radial mass-to-light ratio. 
We also plot the mass interior to the position of the straight arc
in a bi-modal mass model proposed by Pierre \etal (1995) as well as 
display the predictions {}from two simple models for the cluster mass
profile. The solid line shows the prediction for the mass profile 
{}from a singular isothermal sphere with the observed velocity 
dispersion (\cite{carlberg95}), correcting
for the mass contained in the control annulus at each radial bin. 
This mass profile does not appear to agree with the weak lensing mass
profile, predicting little mass for $\theta \geq 150^{\prime\prime}$, 
although uncertainties in the weak lensing estimates are sufficiently large
that the singular model cannot be excluded with a high statistical
significance.  The dashed line corresponds to a model proposed on the 
basis of the CNOC analysis of the cluster galaxy velocity distribution 
(\cite{carlberg96}) under the assumption that the light traces the mass:
$\rho(r) = \rho_0 a / r (r + a)^3$. We have projected the
three-dimensional density model and corrected for the mass in the control 
annulus.  The lensing results and the CNOC model are consistent within the 
95\% CL at all radii probed here.  It is worth noting, however, that
weak lensing profile is rising systematically more steeply than the CNOC
profile for $\theta \geq 150^{\prime\prime}$.

We caution that in the very central regions, the lensing results should
be interpreted with care. First, the analysis performed
here is strictly only valid in the weak lensing regime;  the presence
of the straight arc at a radius of $\simeq 38^{\prime\prime}$ makes this
assumption dubious in the innermost arcminute or so. Secondly, detailed
analyses of other clusters seem to indicate that the lensing signal
in the central regions ($\theta\leq 1^\prime$) is most susceptible to 
dilution due to the contamination of our background galaxy catalogue
by foreground and cluster galaxies.  Since the results presented here
were determined using only $\NgalI$ galaxies, any contamination would 
significantly degrade the lensing signal.  Fortunately, the dilution effect
does not appear to significant beyond a radius of 
$\simeq 1^\prime$. One way to address this is to select only 
galaxies redder than the cluster color sequence -- these should mainly 
lie behind the cluster.
We repeated the above analysis using such a sample and found
consistent results, although the associated uncertainties are
substantially larger since the number of galaxies in the sample 
is much smaller.

Using the cluster light and the lensing mass profiles, we computed the
cluster mass-to-light ratio as a function of radius. 
To account for the subtraction of the material in the control annulus
in the mass estimator, we formed an analogous quantity, $\zeta_L$, for the 
light which is the mean light surface density interior to a given 
radius, relative
to the mean in the control region.  Under the assumption that mass
traces the light --- the cluster light and the lensing mass profiles are
consistent with this assumption, the ratio of the $\zeta$ values based on
the observed shear and $\zeta_L$ forms an unbiased estimate of the 
mass-to-light ratio in the cluster.  The light estimate
comes {}from using all the light associated with the galaxies 
detected in the image (which, as we have 
discussed previously, includes contributions
{}from galaxies not associated with the cluster). We find
a mass-to-light ratio of $M/L_V = (320 \pm 100) h M_\odot/L_{V\odot}$ at
$\theta\approx 2\farcm5 \approx 0.345 \Mpc$, with the radial profile
being consistent with a constant $M/L_V$ across the cluster.

\section{Discussion}
We have mapped the light and mass distributions in the cluster of
galaxies A2390 over a scale of $\simeq 250^{\prime\prime}$ {}from the
cluster center. The projected light
distribution was determined using a color selected galaxy
catalogue. The total radial light profile was calculated using all of 
galaxies detected, which we expect should include some contributions 
{}from galaxies not associated with the cluster. To map the mass distribution,
we measured the shapes of faint galaxies which have been 
gravitationally distorted, correcting for systematic effects in the shape
determinations such as psf anisotropy and seeing.

The projected light and mass distributions are displayed
in Figure \ref{fig:a2390_lightandmass}.
Both the light and mass distributions are centered
on the cD galaxy. The light and mass isocontours are elliptical
in the central region, with the major axis running {}from the southeast
to northwest. There is 
a smaller, secondary concentration in both the cluster galaxy and light 
distributions located approximately $4^\prime$ to the northwest of the
cluster center.
The mass map gives the impression that the mass distribution is extended
in the direction of the 
secondary concentration seen in the cluster light distribution but there is
no corresponding peak in the mass map.  As we noted previously, this is
not surprising given the number of galaxies used in constructing 
the mass map.

Several authors (eg. Pello \etal 1991; Kassiola \etal 1992) have
attempted to account for the lensing features seen in A2390
by suggesting that the cluster has bi-modal mass distribution,
with a small secondary mass concentration located near the observed 
arc position.  This model is further supported by the cluster 
X-ray emission, which shows elliptical isocontours centered on the cluster 
cD galaxy as well as a significant secondary peak
located $40^{\prime\prime}$ to the northwest of the cluster center
(\cite{pierre95}). The weak lensing mass map presented here is not inconsistent
with the notion of a bi-modal mass distribution.  
As we have noted previously, the small-scale
results {}from the weak lensing analysis are subject to greater uncertainties
arising {}from the combined effects of nonlinearity 
in the lensing, contamination by cluster galaxies, and the statistical 
uncertainties due to the small number of galaxies used in the mass 
reconstructions. 

We calculated the mass-to-light ratio as a function of radius 
(Figure \ref{fig:a2390_MassvsR}) and found a
constant value of  $M/L_V = (320 \pm 90) h  M_\odot/L_{\odot V}$,
although we have noted that the mass estimates may be biased downwards
in the innermost $\simeq 1^\prime$ due to contamination by faint
cluster galaxies.
In the course of their study of several intermediate redshift cluster, 
the CNOC group derived a mass-to-light ratio, K-corrected to
redshift zero, of $M/L_r = (331 \pm 54) h  
M_\odot/L_{\odot r}$ for the cluster. We convert the Gunn r luminosity 
into V-band luminosity by applying the transformation
V-r=0.16 (which is typical for a nearby early type galaxy)
and find that the CNOC result corresponds to 
$M/L_V = (370 \pm 60) h  M_\odot/L_{\odot V}$, which is in excellent agreement
with out determination.

We furthermore compared the radial mass profile determined {}from this
analysis with the CNOC model. We projected the CNOC 3D mass model
and calculated the 2D radial mass profile, correcting for mass in
the control annulus to facilitate direct comparison with the
lensing results (Figure \ref{fig:a2390_MassvsR}). 
All of the mass estimates based on the lensing analysis
agree within the 95\% CL with the CNOC model (although we note that
the CNOC model assumes spherical symmetry and there is some indications
that this might be invalid on the scales probed here). At the very
largest radius, there is a slight indication that the lensing inferred
mass might be falling slower than the CNOC model, although the
values are consistent within $2 \sigma$. Larger scale observation of
this cluster will prove useful to determine if this trend continues
at large radii.

As with many of the weak lensing studies performed to date, this analysis
was confined to a relatively small field of view. This has the immediate
consequence that, since the mass determinations made by this technique are
differential, the masses quoted form a {\em lower} bound on the total
mass at any radius. 
The future prospects for this method are exciting as the mass distributions
are mapped out to large distances in the cluster halos.
Indeed, the technology now exists for
such large field lensing observations with the MOCAM 14$^\prime$ 
and the UH $\simeq 0.5^\circ$ cameras at CFHT. 
With the types of
observations possible with these instruments, 
the weak lensing technique offers a unique opportunity to 
probe  many of the outstanding puzzles regarding the dark matter content 
and distribution in the universe.

\end{document}